\documentclass[12pt,twoside]{article}
\usepackage[mathscr]{eucal}
\usepackage{amsmath,amsfonts,amssymb,amsthm,mathabx,empheq}
\usepackage{times}
\usepackage{pdfsync}
\usepackage{cite}
\usepackage{url}
\usepackage{hyperref}
\usepackage{tensor}
\usepackage{color}
\usepackage{multicol}
\usepackage{bbold}
\usepackage[thicklines]{cancel}

\advance\textheight\topskip
\allowdisplaybreaks[1]

\newcommand\cO{{\cal O}}

\newcommand{\p}{\partial}

\newcommand{\be}{\begin{equation}}
\newcommand{\ee}{\end{equation}}
\newcommand{\bea}{\begin{eqnarray}}
\newcommand{\eea}{\end{eqnarray}}

\def \bg {\bar\gamma}

\newcommand{\nn}{\nonumber}
\newcommand*\xbar[1]{%
  \hbox{%
    \vbox{%
      \hrule height 0.5pt 
      \kern0.3ex
      \hbox{%
        \kern-0.0em
        \ensuremath{#1}%
        \kern-0.0em
      }%
    }%
  }%
}

\allowdisplaybreaks[1]

\DeclareFontFamily{OT1}{rsfs}{} \DeclareFontShape{OT1}{rsfs}{m}{n}{
<-7> rsfs5 <7-10> rsfs7 <10-> rsfs10}{}
\DeclareMathAlphabet{\mycal}{OT1}{rsfs}{m}{n}

\hypersetup{
colorlinks=true,
linktoc=page,    
linkcolor=blue,
citecolor=blue,
urlcolor=blue,
colorlinks=true,
}

\begin{document}

\title{From harmonic to Newman-Unti coordinates at the second post-Minkowskian order}

\author{Pujian Mao and Baijun Zeng}
\date{}

\def\mytitle{From harmonic to Newman-Unti coordinates at the second post-Minkowskian order}

\addtolength{\headsep}{4pt}

\begin{centering}

  \vspace{1cm}

  \textbf{\large{\mytitle}}

  \vspace{1cm}

  {\large Pujian Mao and Baijun Zeng}

\vspace{0.5cm}

\begin{minipage}{.9\textwidth}\small \it  \begin{center}
     Center for Joint Quantum Studies, Department of Physics,\\
     School of Science, Tianjin University, 135 Yaguan Road, Tianjin 300350, China
 \end{center}
\end{minipage}

\end{centering}

\begin{center}
Emails:  pjmao@tju.edu.cn,\,\, zeng@tju.edu.cn
\end{center}

\begin{center}
\begin{minipage}{.9\textwidth}
\textsc{Abstract}: In this paper, we present the complete transformations of a generic metric from (generalized) harmonic to Newman-Unti coordinates up to the second post-Minkowskian order $(G^2)$. This allows us to determine the asymptotic shear, the Bondi mass aspect, and the angular-momentum aspect at both orders. 

\end{minipage}
\end{center}

\thispagestyle{empty}


\section{Introduction}

Since the first direct detection of gravitational waves by the LIGO and Virgo collaborations \cite{LIGOScientific:2016aoc}, gravitational-wave physics has been always at the center of attention for many researchers. Although it was predicted by Einstein \cite{Einstein:1916cc,Einstein:1918btx} shortly after the establishment of general relativity, the existence of gravitational waves was still debatable up until the 1960s. The issue is that it was not clear if the gravitational radiation was just an artifact of linearization. The long-standing dispute was resolved in the seminal works of Bondi, van der Burg, Metzner and Sachs \cite{Bondi:1962px,Sachs:1962wk}. They formulated the Bondi–Sachs (BS) formalism \cite{Madler:2016xju}, recasting the Einstein equation near future null infinity as a characteristic initial value problem. Within their framework, gravitational radiation is characterized by the news function and the mass of the gravitational system decreases whenever news function exists. Soon after, the asymptotic structure described by Bondi and collaborators was further clarified in the Newman-Penrose formalism \cite{Newman:1961qr} by Newman and Unti (NU) \cite{Newman:1962cia}.  The BS and NU formulations provide equivalent descriptions of gravitational radiation and can be mapped directly to each other, see, e.g., \cite{Barnich:2011ty}.

One of the advantages of the BS and NU formulations, which are based on asymptotic analysis, is that they provide clear definitions of asymptotically conserved quantities, such as the four-momentum and angular momentum of a radiating gravitational system \cite{Bondi:1962px,Sachs:1962wk,Newman:1965ik,Newman:1968uj,Winicour,Geroch:1977big,Prior,Ashtekar:1978zz,Ashtekar:1979xeo,Ashtekar:1979iaf,Ashtekar:1981bq,Geroch:1981ut,Dray:1984rfa}, see also recent developments in \cite{Barnich:2010eb,Barnich:2011mi,Flanagan:2015pxa}. However, asymptotic analysis can only provide a qualitative characterization of the gravitational wave information at null infinity. The sources that generate these waves in the bulk of spacetime are not captured within the asymptotic framework. Consequently, the precise waveform associated with a given source cannot be obtained directly from either the BS or NU formulations. Quantitative predictions for gravitational waveforms are typically obtained through linearized analyses based on expansions in powers in small parameters, such as the Newton constant $G$, leading to the well established post-Minkowskian (PM) expansions.

The explicit transformation between the PM expansion in harmonic gauge and the asymptotic expansions in BS or NU gauges \cite{Blanchet:2020ngx,Blanchet:2023pce} provides a crucial bridge between the quantitative and qualitative descriptions of gravitational radiation. In particular, expressing PM data within the asymptotic framework greatly clarifies the flux and balance relations associated with gravitational-wave emission \cite{Thorne:1980ru,Blanchet:1985sp,Blanchet:1986dk,Blanchet:1987wq,Damour:1990ji,Blanchet:1992br,Blanchet:1993ng,Blanchet:1996vx,Blanchet:2018yqa,Compere:2019gft},\footnote{In \cite{Bonga:2018gzr}, an inverse transformation from the BS gauge to the harmonic gauge was constructed, providing an alternative route to compute the flux and balance relations.} which play a key role in constructing accurate gravitational-wave templates \cite{Blanchet:2013haa}. The transformations developed in \cite{Blanchet:2020ngx,Blanchet:2023pce} primarily address the multipolar PM approximation. The aim of the present paper is to extend that analysis by providing the complete transformation from harmonic coordinates to NU coordinates for a general perturbative metric up to second post-Minkowskian order $G^2$.

In this paper, we adopt a slightly different strategy to derive the concrete coordinate transformations. In \cite{Blanchet:2020ngx,Blanchet:2023pce}, the transformation between the two coordinate systems was constructed by expressing the NU coordinates in terms of the harmonic coordinates, which manifests the relationship between the two coordinate systems. However, to extract the asymptotic data of the metric in the NU gauge, one must then invert these relations, since the original metric is given in harmonic coordinates. In contrast, we assume in this work that the harmonic coordinates are expressed in terms of the NU coordinates. Furthermore, we assume that the perturbative metric in harmonic coordinates admits an expansion in inverse powers of the radial coordinate at large distances $r$. In standard harmonic coordinates, logarithmic terms arise naturally in the solution of linearized Einstein equation satisfying the harmonic gauge condition. For example, such terms appear at the third post-Newtonian order in the two-body problem \cite{Blanchet:2013haa}. In contrast, we are working with generalized harmonic coordinates under the assumption of a large-distance expansion in integer powers of $1/r$. Nevertheless, logarithmic terms can always be eliminated to arbitrarily high order in $1/r$ expansion by suitably generalizing the harmonic gauge conditions \cite{Duarte:2022mxj}. Under these assumptions, we explicitly compute the coordinate transformations that map a generic perturbative metric into the NU gauge at the first and second PM orders. The asymptotic forms of the perturbative metric in the NU gauge is obtained directly from the change of the coordinates. In particular, we identify the asymptotic data, such as the asymptotic shear, Bondi mass aspect, and angular momentum aspect, from the metric in NU gauge. Our results can be generally applied to any solutions given in the harmonic gauge.

The organization of this paper is as follows. In section \ref{Coor-new}, we present a generic algorithm for deriving the coordinate transformations that connect the harmonic and NU gauges. We derive the NU metric at the linear order in $G$ as a simple illustration of the generic algorithm. The asymptotic shear, Bondi mass aspect, and angular momentum aspect are identified. In section \ref{2PM}, we apply the generic algorithm at the second PM order with respect to the smoothness and stationary conditions at order $G$. The metric in the NU gauge is obtained at $\cO(G^2)$. We conclude in the last section.


\section{Algorithm of the coordinate transformation and the NU metric at linear order}
\label{Coor-new}

We introduce the flat Bondi coordinates $(u_f,r_f,x^A_f)$, which are connected to the harmonic coordinates $(t,x^i)$ as 
\begin{equation}
    t=u_f+r_f,\qquad x^{i}=r_f n^{i}(x^{A}_f),\qquad n^i n_i=1,\qquad (r_f)^2=x^i x_i.
\end{equation}
Correspondingly, the metric components in flat Bondi coordinates are given by
\begin{equation}
\begin{split}
     g^{(f)}_{u_f u_f }&=g_{00}^{(h)},\\
     g^{(f)}_{u_f r_f }&=g_{00}^{(h)} + 2n^{i}g_{0i}^{(h)},\\
     g^{(f)}_{u_f A_f }&=2r_f D_{A_f}n^{i}g_{0i}^{(h)},\\
     g^{(f)}_{r_f r_f }&=g_{00}^{(h)} + 2n^{i}g_{0i}^{(h)} + n^{i}n^{j}g_{ij}^{(h)},\\
     g^{(f)}_{r_f A_f }&=r_f D_{A_f}n^{i}g_{ti}^{(h)} + r_f n^{i}D_{A_f}n^{j}g_{ij}^{(h)},\\
     g^{(f)}_{A_f B_f }&=r_f^2g_{ij}^{(h)} D_{A_f}n^{i}D_{B_f}n^{j},
\end{split}
\end{equation}
where $D_{A_f}$ is the covariant derivative with respect to a unit sphere at the null infinity with metric $\bg_{A_f B_f }$ and $D^2=D^{A_f} D_{A_f}$.

Suppose that the metric is decomposed by $g_{\mu\nu}=\eta_{\mu\nu}+h_{\mu\nu}$, where $\eta_{\mu\nu}$ is the metric of Minkowski spacetime. Asymptotically flat spacetime has the following fall-off behavior $h_{\mu\nu}^{(h)}=\cO(\frac{1}{r_f})$ in harmonic coordinates. Note that $(r_f)^2=x^i x_i$ in harmonic coordinates. The perturbative metric in flat Bondi coordinates can be obtained from those in harmonic coordinates as
\begin{equation}\label{flatmetric}
\begin{split}
     h^{(f)}_{u_f u_f}&=h_{00}^{(h)},\\
     h^{(f)}_{u_f r_f}&=h_{00}^{(h)} + 2n^{i}h_{0i}^{(h)},\\
     h^{(f)}_{u_f A_f}&=2r_f D_{A_f}n^{i} \, h_{0i}^{(h)},\\
     h^{(f)}_{r_f r_f}&=h_{00}^{(h)} + 2n^{i}h_{0i}^{(h)} +n^{i}n^{j}h_{ij}^{(h)},\\
     h^{(f)}_{r_f A_f}&=r_f D_{A_f}n^{i}h_{ti}^{(h)} + r_f n^{i}D_{A_f}n^{j}h_{ij}^{(h)},\\
     h^{(f)}_{A_f B_f}&=r_f^2h_{ij}^{(h)} D_{A_f}n^{i}D_{B_f}n^{j}.
\end{split}
\end{equation}
Correspondingly, one can obtain the asymptotic behavior of the perturbative metric in flat Bondi coordinates
\be
\begin{split}
&h^{(f)}_{u_f u_f}=\cO(\frac{1}{r_f}),\qquad h^{(f)}_{u_f r_f}=\cO(\frac{1}{r_f}), \qquad h^{(f)}_{r_f r_f}=\cO(\frac{1}{r_f}),\\
&h^{(f)}_{u_f A_f}=\cO(1),\qquad h^{(f)}_{r_f A_f}=\cO(1), \qquad h^{(f)}_{A_f B_f}=\cO(r_f).
\end{split}
\ee
Now we introduce the PM expansion for $h_{\mu\nu}$, which is given by
\begin{equation}
h^{(f)}_{\mu\nu}=G {h_1^{(f)}}_{\mu\nu} +G^2 {h_2^{(f)}}_{\mu\nu} + G^3 {h_3^{(f)}}_{\mu\nu} + ....
\end{equation}
The PM expansion for the metric components in flat Bondi coordinates are determined by the metric in harmonic coordinates from \eqref{flatmetric}. Suppose that the perturbative metric in flat Bondi coordinates can be expanded asymptotically near null infinity as
\be\label{expansion}
\begin{split}
&{h_a^{(f)}}_{r_{f} u_{f}}(r_{f},u_{f},x^{A}_{f})=\sum_{i=1}\frac{1}{r_{f}^i}{h_{ai}^{(f)}}_{r_{f} u_{f}}(u_{f},x^{A}_{f}),\\
&{h_a^{(f)}}_{r_{f} r_{f}}(r_{f},u_{f},x^{A}_{f})=\sum_{i=1}\frac{1}{r_{f}^i}{h_{ai}^{(f)}}_{r_{f} r_{f}}(u_{f},x^{A}_{f}) ,\\
&{h_a^{(f)}}_{u_{f} u_{f}}(r_{f},u_{f},x^{A}_{f})=\sum_{i=1}\frac{1}{r_{f}^i}{h_{ai}^{(f)}}_{u_{f} u_{f}}(u_{f},x^{A}_{f}),\\
&{h_a^{(f)}}_{r_f A_f}(r_{f},u_{f},x^{A}_{f})={h_{a0}^{(f)}}_{r_f A_f}(u_{f},x^{A}_{f})+\sum_{i=1}\frac{1}{r_{f}^i}{h_{ai}^{(f)}}_{r_f A_f}(u_{f},x^{A}_{f}),\\
&{h_a^{(f)}}_{u_f A_f}(r_{f},u_{f},x^{A}_{f})={h_{a0}^{(f)}}_{u_f A_f}(u_{f},x^{A}_{f})+\sum_{i=1}\frac{1}{r_{f}^i}{h_{ai}^{(f)}}_{u_f A_f}(u_{f},x^{A}_{f}),\\
&{h_a^{(f)}}_{A_f B_f}(r_{f},u_{f},x^{A}_{f})= r_{f} {h_{am}^{(f)}}_{A_f B_f}(u_{f},x^{A}_{f})+{h_{a0}^{(f)}}_{A_f B_f}(u_{f},x^{A}_{f})\\
&\hspace{6cm}+\sum_{i=1}\frac{1}{r_{f}^i}{h_{ai}^{(f)}}_{A_f B_f}(u_{f},x^{A}_{f}),
\end{split}
\ee
where $a$ indicates the order in the PM expansion. Again the coefficients in the $1/r_f$ expansion for the metric in flat Bondi coordinates can be uniquely determined by the expansions in harmonic coordinates
\be\label{harmonicexpansion}
h_{a\mu\nu}^{(h)}=\sum_{i=1} \frac{h_{ai\mu\nu}^{(h)}}{r_{f}^i},
\ee
at large distance via the relations in \eqref{flatmetric}.

It is important to clarify that we are not working with standard harmonic coordinates, since we do not impose the corresponding harmonic gauge conditions on the perturbative metric. Rather, the coordinates are harmonic only with respect to the flat background. The constraints on the perturbative metric are that it admits a large-distance expansion in integer powers of $1/r_f$. Such conditions can always be achieved within the framework of generalized harmonic coordinates for solutions of linearized Einstein equation \cite{Duarte:2022mxj}.

The equation of motion at linearized level will constrain the coefficient in the large $r_f$ expansion with respect to the stress-energy tensor. Though the precise forms of the stress-energy tensor depend on matter fields. Nevertheless, there are some common features of the asymptotic behaviors for the stress-energy tensor for various matters. Suppose that the stress-energy tensor of matter fields satisfies the following asymptotic behavior
\be\label{stress}
\begin{split}
&T_{r_f r_f}=\cO(r_f^{-4}),\qquad T_{r_f A_f}=\cO(r_f^{-3}),\qquad T_{r_f u_f}=\cO(r_f^{-3}),\\ &T_{u_f u_f}=\cO(r_f^{-2}),\qquad T_{u_f A_f}=\cO(r_f^{-2}),\qquad T_{A_f B_f}=\cO(r_f^{-1}). 
\end{split}
\ee
Note that those conditions are weaker than smoothness of the boundary metric at null infinity, see, e.g., \cite{Satishchandran:2019pyc}. Hence, there will be logarithmic terms in the metric components in the NU gauge. 
In the PM expansion, the stress-energy tensor at higher orders in $G$ (starting at $G^2$) consists of two contributions: one from the matter fields and another from the lower-order perturbative metric. The latter part is referred to as effective stress-energy tensor. The fall-off conditions in \eqref{stress} are intended to hold generally for the full stress-energy tensor at all orders in $G$. This requirement can impose additional constraints on the asymptotic behavior of the perturbative metric, ensuring that the effective stress-energy tensor satisfies the fall-off conditions in \eqref{stress}. These constraints will be discussed explicitly at order $G^2$ at the beginning of the next section. In particular, the effective stress-energy tensor at order $G^2$, which guarantees the smoothness of the boundary metric at null infinity, imposes strong restrictions on the order $G$ metric, which would rule out many physically interesting solutions. Thus we adopt a weaker fall-off conditions for the stress-energy tensor than the smoothness conditions. The constraints from the conditions in \eqref{stress} will be applied to eventually simplify the metric expressions in the NU gauge.\footnote{The algorithm developed in this work can, in principle, be applied without imposing the conditions in \eqref{stress}. However, the resulting metric at order $G^2$ in the NU gauge is very complicated with logarithmic terms appearing in all metric components. For simplicity, we apply the conditions in \eqref{stress} which still encompass most physically interesting systems. } Recalling that the fall-off conditions for the stress-energy tensor is imposed generically, including the effective stress-energy tensor from the metric at lower orders, the linearized Einstein equations at all PM orders should be consistent with those fall-off conditions, which yields that
\be\label{constraints}
\begin{split}
&\p_{u_f} {h_{a1}^{(f)}}_{r_{f} r_{f}}=0,\qquad \p_{u_f} {h_{a0}^{(f)}}_{r_{f} A_f}=0,\qquad \p_{u_f} [\bg^{AB}{h_{am}^{(f)}}_{A_{f} B_{f}}]=0,\\
&{h_{a1}^{(f)}}_{u_{f} r_{f}}=\frac12 {h_{a1}^{(f)}}_{r_{f} r_{f}} - \frac12\p_{u_f} {h_{a2}^{(f)}}_{r_{f} r_{f}}  + \frac14 D^{A_f} D_{A_f} {h_{a1}^{(f)}}_{r_{f} r_{f}},\\
&{h_{a0}^{(f)}}_{u_{f} A_{f}}=\frac12 D_{A_f}  {h_{a1}^{(f)}}_{u_{f} r_{f}} - \frac12 \p_{u_f}  {h_{a1}^{(f)}}_{r_{f} A_{f}} + \frac12 D^{B_f}  {h_{am}^{(f)}}_{A_{f} B_{f}}\\
&\qquad -\frac12 D_{A_f} (\bg^{{C_f} D_f}{h_{am}^{(f)}}_{C_{f} D_{f}}) + \frac12 D^2 {h_{a0}^{(f)}}_{r_{f} A_{f}} - \frac12 D^{B_f} D_{A_f} {h_{a0}^{(f)}}_{r_{f} B_{f}} \\
&\qquad + \frac14D_{A_f} D^2 {h_{a1}^{(f)}}_{r_{f} r_{f}} +{h_{a0}^{(f)}}_{r_{f} A_{f}}.
\end{split}
\ee

We assume that the transformations from flat Bondi coordinates to the NU coordinates $(u,r,x^A)$ can also be given in the PM expansion as
\begin{equation}
\begin{split}
    &u_{f}=u +GU_{1}(u,r,x^A)+G^{2}U_{2}(u,r,x^A)+G^3U_{3}(u,r,x^A)+...,\\
    &r_{f}=r +G R_{1}(u,r,x^A)+G^{2}R_{2}(u,r,x^A)+G^3R_{3}(u,r,x^A)+...,\\
    &x^{A}_{f}=x^{A} +G X^{A}_{1}(u,r,x^A)+G^{2}X^{A}_{2}(u,r,x^A)+G^3X^{A}_{3}(u,r,x^A)+... .
    \label{transformation-new}
\end{split}
\end{equation}
The gauge conditions of the NU framework are
\begin{equation}\label{NU}
    g_{rr}=0,\quad g_{ru}=-1, \quad g_{rA}=0 .
\end{equation}
The strategy to construct the perturbative diffeomorphism in $G$ expansion is as follows. Starting from the metric in flat Bondi coordinates, the transformed metric is obtained as
\be\label{transf}
g_{\mu\nu}= g^{(f)}_{\alpha\beta} \frac{\p x_f^\alpha}{\p x^\mu}\frac{\p x_f^\beta}{\p x^\nu}.
\ee
The NU gauge conditions yield the following transformation laws
\be
g^{(f)}_{\alpha\beta} \frac{\p x_f^\alpha}{\p r}\frac{\p x_f^\beta}{\p r}=0,\qquad g^{(f)}_{\alpha\beta} \frac{\p x_f^\alpha}{\p u}\frac{\p x_f^\beta}{\p r}=-1,\qquad g^{(f)}_{\alpha\beta} \frac{\p x_f^\alpha}{\p x^A}\frac{\p x_f^\beta}{\p r}=0.
\ee
Inserting the relations of the two coordinate systems \eqref{transformation-new}, one can derive  $U_{1},R_{1},X^{a}_{1}$ at 1PM as
\begin{align}
    &\partial_r U_{1}=\frac{1}{2} {h_1^{(f)}}_{r_f r_f}, \label{U1}\\
    &\partial_r R_{1}={h_1^{(f)}}_{u_f r_f} - \frac12 {h_1^{(f)}}_{r_f r_f} - \p_u U_1, \label{R1}\\
    &\partial_r X^{A}_{1}=\frac{1}{r^2}D^A U_1 - \frac{1}{r^2}\bar\gamma^{AB} {{h_1^{(f)}}_{r_f B}}, \label{X1}
\end{align}
where $\bar\gamma^{AB}$ is the inverse metric of the celestial sphere in NU coordinates. Since the perturbative metric $h_{\mu\nu}$ satisfies asymptotically flat conditions, the celestial sphere metric has the same form for flat Bondi and NU coordinates. Moreover, the difference between flat Bondi coordinates and NU coordinates starts at order $G$, namely $x_f^A=x^A + \cO(G)$. We will not distinguish the index $A_f$ with $A$ at a fixed PM order as they represent exactly the same information. The capital Latin index will be raised and lowered by the celestial sphere metric. With respect to the asymptotic expansions \eqref{expansion}, equations \eqref{U1}-\eqref{X1} can be solved as
\begin{align}
&U_1=\frac{1}{2}{h_{11}^{(f)}}_{r_f r_f} \log r +  U_{10}(u,x^A) - \frac{{h_{12}^{(f)}}_{r_f r_f} }{2r}  - \frac{{h_{13}^{(f)}}_{r_f r_f} }{4r^2} + \cO(\frac{1}{r^3}), \label{UG}\\
&R_1= - \p_u U_{10} r - \left[ \frac12 {h_{11}^{(f)}}_{r_f r_f} -{h_{11}^{(f)}}_{u_f r_f} - \frac12 \p_u {h_{12}^{(f)}}_{r_f r_f} \right] \log r \label{RG}\\
&\qquad + R_{10}(u,x^A) + \frac{2{h_{12}^{(f)}}_{r_f r_f} - 4{h_{12}^{(f)}}_{u_f r_f} - \p_u {h_{13}^{(f)}}_{r_f r_f} }{4r} +\cO(\frac{1}{r^2}), \nn\\
&X^A_1=X_{10}^A(u,x^A) - \frac{D^A {h_{11}^{(f)}}_{r_f r_f} \log r}{2r} + \frac{2{{h_{10}^{(f)}}_{r_f}}^{A}  - D^A {h_{11}^{(f)}}_{r_f r_f} - 2 D^A U_{10}}{2r}\nn \\
&\qquad +\frac{2 {{h_{11}^{(f)}}_{r_f}}^{A}   +D^A {h_{12}^{(f)}}_{r_f r_f} }{4r^2}  +\frac{4 {{h_{12}^{(f)}}_{r_f}}^{A}  +D^A {h_{13}^{(f)}}_{r_f r_f} }{12r^3} + \cO(\frac{1}{r^4}), \label{XG}
\end{align}
where $U_{10}$, $R_{10}$, and $X_{10}^A$ are integration constants with respect to the radial variable $r$. Above, we only present the orders that are relevant to asymptotic data in the NU framework in the $1/r$ expansion. The fall-off conditions of the NU gauge are \cite{Newman:1962cia,Barnich:2011ty}
\be\label{fall-off}
g_{uA}=o(r),\,\,g_{uu}=-1 + o(1),\,\, g_{AB}=r^2  \bar\gamma_{A B}  + o(r^2),\,\,
|g_{AB}|=r^4|\bg_{AB}|+o(r^3) .
\ee
Those conditions fix the four integration constants in the NU coordinates as
\begin{align}
&X_{10}^A=Y_1^A(x^B), \\
&U_{10}=  \beta_1(x^A) + u D_A Y_1^A, \\
&R_{10}=-\frac14 \bar{\gamma}^{AB}{h_{1m}^{(f)}}_{A_{f} B_{f}} + \frac14 D^A D_A {h_{11}^{(f)}}_{r_f r_f} - \frac12 D^A {h_{10}^{(f)}}_{r_f A}  + \frac12 D^2 U_{10},
\end{align}
where $Y_1^A$ represents conformal Killing vector on the celestial sphere at order $G$. From the perspective of a full diffeomorphism transformation, $Y_1^A$ generate Lorentz transformations. Since these are isometries of the Minkowski spacetime, they do not affect the perturbative metric in the NU gauge. We have verified this explicitly through direct computation. Hence, we can set $Y_1^A=0$ for simplicity without altering the final metric in NU gauge. On the other hand, $\beta_1$ represents the supertranslation at order $G$.

In this work, we apply a slightly different strategy than \cite{Blanchet:2020ngx,Blanchet:2023pce} for deriving the metric in the NU gauge. Specifically, we express the flat Bondi coordinates in terms of the NU coordinates. Hence, the remaining metric components can be obtained directly by inserting the relations in \eqref{transformation-new} into \eqref{transf} with the order $G$ data given in \eqref{UG}-\eqref{XG}. Finally, the components of the perturbative metric at order $G$ in the NU gauge are obtained as
\begin{align}
&{h_1}_{uu}= \frac{1}{r}\left[ {h_{11}^{(f)}}_{u_f u_f}  + 2 \p_u {h_{12}^{(f)}}_{u_f r_f}  + \frac12 \p_u^2 {h_{13}^{(f)}}_{r_f r_f} \right] + \cO(\frac{1}{r^2}),\\
&{h_1}_{AB}=   \tilde{{C_1}}_{AB} r \log r +  {C_1}_{AB} r + \cO(1), \\
&\qquad \tilde{{C_1}}_{AB} = -D_A D_B {h_{11}^{(f)}}_{r_f r_f} + \frac12 \bg_{AB} D^2 {h_{11}^{(f)}}_{r_f r_f} , \\
&\qquad {C_1}_{AB}= {h_{1m}^{(f)}}_{A  B } + 2 D_{(A} {h_{10}^{(f)}}_{B) r_f} + \tilde{{C_1}}_{AB} - 2 D_A D_B \beta_1 + \bg_{AB}D^2\beta_1 \nn\\
&\qquad \qquad \qquad  - \frac12 \bg_{AB} \bg^{CD}{h_{1m}^{(f)}}_{C  D } - \bg_{AB} D^C {h_{10}^{(f)}}_{r_f C} , \\
&{h_1}_{uA}= \bigg(\frac12 D^B \tilde{{C_1}}_{AB} \log r + \frac12 D^B {{C_1}}_{AB} - \frac34 D^B  \tilde{{C_1}}_{AB} \bigg) + \frac{1}{r} \bigg[ {h_{11}^{(f)}}_{u_f A}  + D_A {h_{12}^{(f)}}_{u_f r_f} \nn\\
&\qquad \qquad  + \frac13 \p_u {h_{12}^{(f)}}_{r_f A} + \frac13 \p_u D_A {h_{13}^{(f)}}_{r_f r_f} \bigg]  + \cO(\frac{1}{r^2}).
\end{align}
The extra term $- \frac34 D^B  \tilde{C}_{AB}$ in $g_{uA}$ than the standard NU solution space is precisely from the logarithmic terms in $g_{AB}$. Note that the constraints in \eqref{constraints} have been applied. We have verified the above generic results for a special
case of a system of pointlike bodies source. The final results, when further transformed into the Bondi gauge following \cite{Barnich:2011ty}, recover precisely that in \cite{Mao:2024urq}. 

To conclude this section, we briefly comment on the order $G$ metric in NU gauge. The logarithmic terms are only relevant to ${h_{11}^{(f)}}_{r_f r_f}$, which fully aligns with the smoothness conditions in \cite{Satishchandran:2019pyc}. The shear tensor consists of both ${h_{1m}^{(f)}}_{A  B }$ and ${h_{10}^{(f)}}_{A r_f}$ components and the supertranslation $\beta_1$. But the news tensor
\be
{N_1}_{AB}=\p_u {C_1}_{AB}=\p_{u_f} {h_{1m}^{(f)}}_{A  B }  - \frac12 \bg_{AB} \bg^{CD} \p_{u_f} {h_{1m}^{(f)}}_{C  D } ,
\ee
is completely fixed by the traceless part of the linear transverse metric in the flat Bondi coordinates, applying the constraints in \eqref{constraints}.
An interesting fact is that the mass and angular momentum aspects are determined by data from three different orders in the $1/r_f$ expansion in the flat Bondi coordinates. More precisely,
\be
m_1= {h_{11}^{(f)}}_{u_f u_f}  + 2 \p_u {h_{12}^{(f)}}_{u_f r_f}  + \frac12 \p_u^2 {h_{13}^{(f)}}_{r_f r_f} ,
\ee
and\footnote{In the standard Bondi or NU gauge, the angular momentum aspect is not the full $g_{uA}$ component at order $1/r$, see, e.g.,  \cite{Barnich:2010eb,Flanagan:2015pxa}. However, the difference is only a total derivative term which does not affect the definition of angular momentum. In this work, we only introduce some fall-off conditions for the stress-energy tensor without specifying its precise formulas. Hence, the perturbative metric is only a solution to the linearized Einstein equation to the orders with respect to the fall-off conditions of the stress-energy tensor in $\frac1r$ expansion. The exact expressions of the obtained metric components in NU gauge may not fully align with the results in \cite{Barnich:2010eb,Flanagan:2015pxa} where the metric is a solution of Einstein equation to all orders. }
\be
{N_1}_A= {h_{11}^{(f)}}_{u_f A}  + D_A {h_{12}^{(f)}}_{u_f r_f}  + \frac13 \p_u {h_{12}^{(f)}}_{r_f A} + \frac13 \p_u D_A {h_{13}^{(f)}}_{r_f r_f}.
\ee
Nevertheless, the contributions from subleading orders arise only from either the evolution or the divergence on the celestial sphere of the metric components at those subleading orders. The mass and angular momentum aspects are independent of supertranslations.


\section{Metric in NU gauge at quadratic order}
\label{2PM}

In this section, we will derive the metric in NU gauge at order $G^2$. Before computing the coordinate transformations, we first revisit the fall-off conditions for the stress-energy tensor imposed in \eqref{stress}. At order $G^2$, the stress-energy tensor contains two types of contribution, namely, the matter stress–energy tensor and the effective stress–energy tensor induced by the order $G$ perturbative metric. We find that the effective stress–energy tensor constructed from the general asymptotic form of the perturbative metric in \eqref{expansion} does not fully obey the fall-off conditions in \eqref{stress}. Therefore, for a consistent computation at order $G^2$, we must impose stronger fall-off conditions at $\cO(G)$, namely,\footnote{The algorithm developed in this work can be definitely applied to the computation at $\cO(G^2)$ without imposing these extra conditions. Correspondingly, the relations in \eqref{constraints} are not valid at $\cO(G^2)$. However, imposing those conditions can lead to a consistent computation for both PM orders, which could manifest some generic features at different PM orders.}
\be\label{more}
{h_{11}}_{r_f r_f}=0, \qquad \p_u  {h_{1m}}_{A_f B_f}=0, \qquad \p_u  {h_{11}}_{r_f u_f}=0.
\ee
The first condition requires the metric at order $G$ to be smooth, i.e., free of logarithmic terms. This condition is typically related to the choice of reference system. For instance, it corresponds to the center-of-mass reference for a system of pointlike bodies source \cite{Veneziano:2022zwh}. The second condition implies that there is no news at order $G$, meaning that the linear order metric is stationary. This situation is generic, e.g., in the classical gravitational scattering \cite{Damour:2020tta}. The last condition can be achieved by an appropriate gauge transformation in flat Bondi coordinates, which is given by 
\be
\chi_\mu=\left(0,\frac{\chi_{r_f}}{r_f},0,0\right).
\ee
The perturbative metric components are  transformed as
\be
{h_{12}'}_{r_f r_f}={h_{12}}_{r_f r_f}-2\chi_r,\,\, {h_{11}'}_{u_f r_f}={h_{11}}_{u_f r_f}+\p_{u_f}\chi_r,\, \,{h_{11}'}_{A_f r_f}={h_{11}}_{A_f r_f}+\p_{A_f}\chi_r.
\ee
This transformation allows us to set $\p_u  {h_{11}}_{r_f u_f}=0$ without affecting any of the previously imposed fall-off conditions in the flat Bondi coordinates. We will now continue to compute the order $G^2$ metric in the NU gauge subject also to the additional requirements given in \eqref{more}.

The NU gauge conditions \eqref{NU} determine $U_{2},R_{2},X^{A}_{2}$ in \eqref{transformation-new} at 2PM as
\begin{align}
\partial_r U_{2}=&\frac{1}{2}{h^{(f)}_{2}}_{r_{f}r_{f}} 
-\frac{1}{2} (\partial_{r} U_1)^2 -  \partial_{r} U_1 \partial_{r}R_1 + \frac{1}{2} r^2 \bg_{AB}\partial_{r} X_{1}^{A}\partial_{r}X_{1}^{B}  \nn\\
&+ {h^{(f)}_{1}}_{r_{f}u_f}\partial_{r} U_1 + {h^{(f)}_{1}}_{r_{f}r_f}\partial_{r} R_1    + {h^{(f)}_{1}}_{r_{f}A}\partial_{r} X^A_1  + \frac12 U_1  \p_u {h^{(f)}_{1}}_{r_{f}r_{f}} \nn\\
&+ \frac12 R_1  \p_r {h^{(f)}_{1}}_{r_{f}r_{f}}+ \frac12 X^A_1  D_A {h^{(f)}_{1}}_{r_{f}r_{f}}, \label{U2}\\
\partial_r R_{2}=&{h^{(f)}_{2}}_{u_{f}r_{f}}  -\partial_{r}U_1 \partial_{u}U_1 - \partial_{r}U_1 \partial_{u} R_1 - \partial_{r}R_1 \partial_{u}U_1 + r^2 \bg_{AB}\partial_{r}X_{1}^A \partial_{u}X_{1}^B \nn \\
& +{h^{(f)}_{1}}_{r_{f}u_f}\partial_{u}U_1   +{h^{(f)}_{1}}_{r_{f}r_f}\partial_{u}R_1 +{h^{(f)}_{1}}_{r_{f}A}\partial_{u}X_{1}^{A} +{h^{(f)}_{1}}_{u_{f}u_f}\partial_{r} U_1  \nn\\
&  +{h^{(f)}_{1}}_{u_{f} r_f }\partial_{r}R_1 +{h^{(f)}_{1}}_{u_{f}A}\partial_{r}X_{1}^{A}  +  U_1  \p_u {h^{(f)}_{1}}_{u_{f}r_{f}} +  R_1  \p_r {h^{(f)}_{1}}_{u_{f}r_{f}} \nn\\
& +  X_1^A  D_A {h^{(f)}_{1}}_{u_{f}r_{f}} -\partial_r U_{2}-\partial_{u}U_{2} , \label{R2}\\
\partial_r X^{A}_{2}=&-\frac{\bg^{AB}}{r^2}\bigg( {h^{(f)}_{2}}_{r_{f}B} +{h^{(f)}_{1}}_{r_{f} u_f } D_B U_{1} +{h^{(f)}_{1}}_{r_{f}r_f} D_B R_{1} +{h^{(f)}_{1}}_{r_{f} C} D_B X_{1}^{C} \nn\\
&  + U_{1} \partial_u {h^{(f)}_{1}}_{{r_{f}B}} + R_{1} \partial_r {h^{(f)}_{1}}_{{r_{f}B}}+ X_{1}^{C} D_C {h^{(f)}_{1}}_{{r_{f}B}} + {h^{(f)}_{1}}_{B u_f}\partial_{r} U_{1} \nn\\
&  + {h^{(f)}_{1}}_{B r_f}\partial_{r}R_{1}  + {h^{(f)}_{1}}_{B C}\partial_{r}X_{1}^{C} - \partial_{r} U_{1} D_B U_{1} - \partial_{r} U_{1} D_B R_{1}  - \partial_{r} R_{1} D_B U_{1} \nn\\
& - D_B U_2 + 2r R_1 \p_r X_{1B} \bigg) - \bg^{AB}   \partial_{r}X_{1C} \p_B X_{1}^{C} + X_{1}^{C}\partial_{C}\bg^{AB}\partial_{r}X_{1B}. \label{X2}
\end{align}
A noteworthy feature of the last two terms in the above equation is that $X_2^A$ does not transform as a vector on the celestial sphere. The origin of this non-tensorial behavior is the $\cO(G)$ correction to the celestial sphere metric that appears when transforming from flat Bondi coordinates to NU coordinates, given by $G X_{1}^{C}\partial_{C}\bg^{AB}$. This term enters precisely at the order $G^2$ computations. Nevertheless, this is consistent with the full BMS transformation studied in \cite{Flanagan:2023jio} where non-tensorial terms also arise at subleading orders in the angular coordinate transformations. Using the expansion in \eqref{expansion} together with the order $G$ coordinates in \eqref{UG}-\eqref{XG}, we obtain the order $G^2$ coordinates up to integration constants as
\begin{equation}
\begin{split}
U_2=&\frac{1}{2}{h_{21}^{(f)}}_{r_f r_f}  \log r +  U_{20}(u,x^A) - \frac{1}{2r} \bigg[ {h_{22}^{(f)}}_{r_f r_f} - {h_{10}^{(f)}}_{r_f A_f} {{h_{10}^{(f)}}_{r_f}}^{A_f}\\
&+ \beta_1 \p_u {h_{12}^{(f)}}_{r_f r_f} + D^A \beta_1 D_A \beta_1
\bigg] + \frac{1}{4r^2} \bigg[-{h_{23}^{(f)}}_{r_f r_f}  + 2 {h_{11}^{(f)}}_{r_f A_f}{{h_{10}^{(f)}}_{r_f}}^{ A_f} \\
&  - \frac12 {h_{12}^{(f)}}_{r_f r_f} {{h_{1m}^{(f)}}_{A_f}}^{ A_f} - D^A({h_{12}^{(f)}}_{r_f r_f}  {h_{10}^{(f)}}_{r_f A})+ \frac12 \p_u ({h_{12}^{(f)}}_{r_f r_f})^2  \\
& + {h_{12}^{(f)}}_{r_f r_f} D^2 \beta_1 + 2 D^A \beta_1 D_A {h_{12}^{(f)}}_{r_f r_f} - \beta_1 \p_u {h_{13}^{(f)}}_{r_f r_f}\bigg]+ \cO(\frac{1}{r^3}),
\end{split}
\end{equation}
\begin{align}
R_2=&  -  \p_u  U_{20} r - \left[ \frac12 {h_{21}^{(f)}}_{r_f r_f} -{h_{21}^{(f)}}_{u_f r_f} - \frac12 \p_u {h_{22}^{(f)}}_{r_f r_f} \right] \log r + R_{20}(u,x^A)\nn \\
& + \frac{2{h_{22}^{(f)}}_{r_f r_f} - 4{h_{22}^{(f)}}_{u_f r_f} - \p_u {h_{23}^{(f)}}_{r_f r_f} }{4r} 
+\frac{\beta_1 \p_u\left(2{h_{12}^{(f)}}_{r_f r_f} - 4{h_{12}^{(f)}}_{u_f r_f} - \p_u {h_{13}^{(f)}}_{r_f r_f} \right)}{4r} \nn \\
&+ \frac{1}{2r} \bigg[ {h_{10}^{(f)}}_{r_f A_f} D_B {h_{1m}^{(f)}}^{A_f B_f} -  {h_{10}^{(f)}}_{r_f A} D^A {{h_{1m}^{(f)}}_{B_f}}^{B_f} + {{h_{10}^{(f)}}_{r_f}}^{A_f} D^2 {h_{10}^{(f)}}_{r_f A_f}  \\
&+ {h_{10}^{(f)}}_{r_f A_f} {{h_{10}^{(f)}}_{r_f}}^{A_f} -  {h_{10}^{(f)}}_{r_f A} D^B D^A {h_{10}^{(f)}}_{r_f B} + D^A \beta_1 D_A {{h_{1m}^{(f)}}_{B_f}}^{B_f} - D^A \beta_1  D^2 {h_{10}^{(f)}}_{r_f A} \nn \\
&-2 D^A \beta_1 {h_{10}^{(f)}}_{r_f A}- D^A \beta_1 D^B {h_{1m}^{(f)}}_{AB}  +D^A \beta_1  D_B  D_A {{h_{10}^{(f)}}_{r_f}}^{ B} 
 + D^A \beta_1 D_A \beta_1 \bigg]+\cO(\frac{1}{r^2}),\nn
\end{align}
\begin{align}
X^A_2=&X_{20}^A(u,x^A) - \frac{D^A {h_{21}^{(f)}}_{r_f r_f} \log r}{2r}
+ \frac{2{{h_{20}^{(f)}}_{r_f}}^{A_f} (x^A) - D^A {h_{21}^{(f)}}_{r_f r_f}(x^A) - 2 D^A U_{20}}{2r} \nn \\
&+ \frac{1}{2r^2} \bigg[ {{h_{21}^{(f)}}_{r_f}}^{A} +  {h_{11}^{(f)}}_{u_f r_f} D^A \beta_1 -{{h_{1m}^{(f)}}_{B}}^{A}\left({{h_{10}^{(f)}}_{r_f}}^{B} - D^B \beta_1\right) \nn \\   
&\qquad -2 R_{10} \left({{h_{10}^{(f)}}_{r_f}}^{A} - D^A \beta_1\right) - D^A U_{21} + \beta_1 \p_u {{h_{11}^{(f)}}_{r_f}}^{A}  \nn\\
&\qquad + \left({{h_{10}^{(f)}}_{r_f}}^{B} - D^B \beta_1\right) D_B {{h_{10}^{(f)}}_{r_f}}^A + {{h_{10}^{(f)}}_{r_f}}^{B} D^A \left({{h_{10}^{(f)}}_{r_f B}}  - D_B \beta_1\right) \nn \\
& \qquad - \bg^{AB}  \left({{h_{10}^{(f)}}_{r_f C}}  - D_C \beta_1\right) {\p_B} \left({{h_{10}^{(f)}}_{r_f}}^{C} - D^C \beta_1\right) \nn\\
&\qquad + \left({{h_{10}^{(f)}}_{r_f B}}  - D_B \beta_1\right)\left({{h_{10}^{(f)}}_{r_f}}^{C} - D^C \beta_1\right) \p_C \bg ^{AB}\bigg] \nn \\
&+\frac{1}{3r^3} \bigg[  {{h_{22}^{(f)}}_{r_f}}^{A} - \frac12 {h_{11}^{(f)}}_{u_f r_f} D^A {h_{12}^{(f)}}_{r_f r_f}- \frac12 {h_{12}^{(f)}}_{r_f r_f} \p_u {{h_{11}^{(f)}}_{r_f}}^A +\frac12 {{h_{10}^{(f)}}_{u_f}}^A {h_{12}^{(f)}}_{r_f r_f} \nn \\
&\qquad - \frac32 {{h_{10}^{(f)}}_{r_f}}^A \left( {h_{12}^{(f)}}_{r_f r_f} - 2{h_{12}^{(f)}}_{u_f r_f} -\frac12\p_u {h_{13}^{(f)}}_{r_f r_f}\right) \\
&\qquad - {{h_{10}^{(f)}}_{B}}^A 
\left( {{h_{10}^{(f)}}_{r_f}}^B - D^B \beta_1\right) - {{h_{1m}^{(f)}}_{B}}^A 
\left( {{h_{11}^{(f)}}_{r_f}}^B + \frac12 D^B {h_{12}^{(f)}}_{r_f r_f}\right) \nn \\
&\qquad + \frac12 {{h_{10}^{(f)}}_{r_f}}^B D^A 
\left( {{h_{11}^{(f)}}_{r_f B}} + \frac12 D_B {h_{12}^{(f)}}_{r_f r_f}\right) + {{h_{11}^{(f)}}_{r_f B}} D^A \left( {{h_{10}^{(f)}}_{r_f}}^B - D^B \beta_1\right) \nn \\
&\qquad + \frac12 \left({{h_{11}^{(f)}}_{r_f}}^B + \frac12 D^B  {h_{12}^{(f)}}_{r_f r_f}\right) 
 D_B {{h_{10}^{(f)}}_{r_f}}^A + \left( {{h_{10}^{(f)}}_{r_f}}^B - D^B \beta_1\right) D_B {{h_{11}^{(f)}}_{r_f}}^A \nn \\ 
&\qquad +  D^A \beta_1 \left( {h_{12}^{(f)}}_{r_f r_f} - 2 {h_{12}^{(f)}}_{u_f r_f} - \frac34\p_u {h_{13}^{(f)}}_{r_f r_f}\right)  + \frac12{h_{12}^{(f)}}_{r_f r_f} D^A R_{10}  \nn\\  
&\qquad -  R_{10} \left( 3{{h_{11}^{(f)}}_{r_f}}^A +  D^A {h_{12}^{(f)}}_{r_f r_f}\right) - D^A U_{22} + \beta_1 \p_u {{h_{12}^{(f)}}_{r_f}}^A \nn \\
&\qquad - \bg^{AB}  \left({{h_{11}^{(f)}}_{r_f C}} + \frac12 D_C  {h_{12}^{(f)}}_{r_f r_f}\right)  \p_B \left({{h_{10}^{(f)}}_{r_f}}^C - D^C \beta_1 \right) \nn\\
&\qquad -\frac12 \bg^{AB} \left({{h_{10}^{(f)}}_{r_f C }} - D_C \beta_1 \right)  \p_B \left({{h_{11}^{(f)}}_{r_f }}^C + \frac12 D^C  {h_{12}^{(f)}}_{r_f r_f}\right) \nn \\
&\qquad +\p_C \bg^{AB} \left({{h_{11}^{(f)}}_{r_f B}} + \frac12 D_B  {h_{12}^{(f)}}_{r_f r_f}\right)  \left({{h_{10}^{(f)}}_{r_f}}^C - D^C \beta_1 \right) \nn\\
&\qquad + \frac12 \p_C \bg^{AB} \left({{h_{10}^{(f)}}_{r_f B }} - D_B \beta_1 \right) \left({{h_{11}^{(f)}}_{r_f }}^C + \frac12 D^C  {h_{12}^{(f)}}_{r_f r_f}\right) \bigg]
+ \cO(\frac{1}{r^4}),\nn
\end{align}
where $U_{21}$ and $U_{22}$ are the coefficients of $U_2$ at order $\frac1r$ and $\frac{1}{r^2}$ respectively. The fall-off conditions \eqref{fall-off} fix the integration constants as
\begin{align}
&X_{20}^A=Y_2^A(x^B), \\
&U_{20}=  \beta_2(x^A) + u D_A Y_2^A, \\
&R_{20}=-\frac14 \bar{\gamma}^{AB}{h_{2m}^{(f)}}_{A_{f} B_{f}} + \frac14 D^A D_A {h_{21}^{(f)}}_{r_f r_f} - \frac12 D^A {h_{20}^{(f)}}_{r_f A}  + \frac12 D^2 U_{20} ,
\end{align}
where $Y_2^A$ represents conformal Killing vector on the celestial sphere at order $G^2$, which can be fixed as $Y_2^A=0$ without affecting the final metric in NU gauge as discussed in previous section. $\beta_2$ is the supertranslation at order $G^2$.

Finally, the components of the perturbative metric at order $G^2$ in the NU gauge are obtained as
\begin{align}
&{h_2}_{uu}= \frac{1}{r}\bigg[ {h_{21}^{(f)}}_{u_f u_f}  + 2 \p_u {h_{22}^{(f)}}_{u_f r_f}  + \frac12 \p_u^2 {h_{23}^{(f)}}_{r_f r_f}  \\
&\qquad\qquad + \beta_1 \p_u \left( {h_{11}^{(f)}}_{u_f u_f}  + 2 \p_u {h_{12}^{(f)}}_{u_f r_f}  + \frac12 \p_u^2 {h_{13}^{(f)}}_{r_f r_f}\right)\bigg] + \cO(\frac{1}{r^2}),\nn\\
&{h_2}_{AB}=  \tilde{{C_2}}_{AB} r \log r + {C_2}_{AB} r + \cO(1), \\
&\qquad \tilde{{C_2}}_{AB} =   -D_A D_B {h_{21}^{(f)}}_{r_f r_f} + \frac12 \bg_{AB} D^2 {h_{21}^{(f)}}_{r_f r_f}, \\
&\qquad {C_2}_{AB}= {h_{2m}^{(f)}}_{A  B } + 2 D_{(A} {h_{20}^{(f)}}_{B) r_f}  -D_A D_B {h_{21}^{(f)}}_{r_f r_f}  - 2 D_A D_B \beta_2  + \bg_{AB}D^2\beta_2 \nn \\
&\qquad \qquad \qquad  - \frac12 \bg_{AB} \bg^{CD}{h_{2m}^{(f)}}_{C  D } - \bg_{AB} D^C {h_{20}^{(f)}}_{r_f C} + \frac12 \bg_{AB} D^2 {h_{21}^{(f)}}_{r_f r_f}, \\
&{h_2}_{uA}= \bigg(\frac12 D^B \tilde{{C_2}}_{AB} \log r + \frac12 D^B {{C_2}}_{AB} - \frac34 D^B  \tilde{{C_2}}_{AB} \bigg) + \frac{{h_{21}}_{uA}}{r}  + \cO(\frac{1}{r^2}),
\end{align}
where 
\begin{align}
  {h_{21}}_{uA}= & {h_{21}^{(f)}}_{u_f A}  + D_A {h_{22}^{(f)}}_{u_f r_f}+ \frac13 \p_u {h_{22}^{(f)}}_{r_f A} + \frac13 \p_u D_A {h_{23}^{(f)}}_{r_f r_f} \nn \\
& +\beta_1 \p_u\left( {h_{11}^{(f)}}_{u_f A}  + D_A {h_{12}^{(f)}}_{u_f r_f}+ \frac13 \p_u {h_{12}^{(f)}}_{r_f A} + \frac13 \p_u D_A {h_{13}^{(f)}}_{r_f r_f}\right) \nn\\
& + \frac13 D_A \beta_1 \left( 3{h_{11}^{(f)}}_{u_f u_f}  + 4 \p_u {h_{12}^{(f)}}_{u_f r_f}  +  \p_u^2 {h_{13}^{(f)}}_{r_f r_f} + \p_u {h_{12}^{(f)}}_{r_f r_f}\right) \nn \\
&  + \frac13 D^B \beta_1 \p_u \left(  {h_{10}^{(f)}}_{AB}  +  D_{(A}{h_{10}^{(f)}}_{B) r_f} + \frac12 D_A D_B  {h_{12}^{(f)}}_{r_f r_f} \right) \nn\\
&  + D^B \beta_1 \left( D_A {h_{10}^{(f)}}_{r_f B} - D_B {h_{10}^{(f)}}_{r_f A} \right)+ \frac12 D^B \beta_1 D^2 \left( D_A {h_{10}^{(f)}}_{r_f B} - D_B {h_{10}^{(f)}}_{r_f A} \right) \nn \\
&  + \frac12 D^B \beta_1 D^C \left( D_A {h_{1m}^{(f)}}_{B C} - D_B {h_{1m}^{(f)}}_{A C} \right) \\
&    - \frac13 {h_{10}^{(f)}}_{r_f A} \p_u {h_{12}^{(f)}}_{r_f r_f} 
  - \frac{1}{6} {h_{10}^{(f)}}_{r_f B} D_A D^B \p_u {h_{12}^{(f)}}_{r_f r_f}+ \frac{1}{12} {h_{1m}^{(f)}}_{A B} D^B \p_u {h_{12}^{(f)}}_{r_f r_f} \nn \\ 
&  + \frac16 \p_u {h_{12}^{(f)}}_{r_f r_f} \left(D_B D_A {{h_{10}^{(f)}}_{r_f}}^B  - D^2  {h_{10}^{(f)}}_{r_f A} + D_A {{h_{1m}^{(f)}}_{B_f}}^{B_f} - D^B {h_{1m}^{(f)}}_{A B}\right) \nn \\
&  + \frac{1}{12} \left(D^B  {h_{10}^{(f)}}_{r_f A} D_B \p_u {h_{12}^{(f)}}_{r_f r_f} - D_A  {{h_{10}^{(f)}}_{r_f}}^B D_B \p_u {h_{12}^{(f)}}_{r_f r_f} \right) \nn \\
&    - \frac12 {{h_{10}^{(f)}}_{r_f}}^B  D^C \left( D_A {h_{1m}^{(f)}}_{B C} - D_B {h_{1m}^{(f)}}_{A C} \right)  \nn \\
&    -\frac12{{h_{10}^{(f)}}_{r_f}}^B D^2 \left( D_A {h_{10}^{(f)}}_{r_f B} - D_B {h_{10}^{(f)}}_{r_f A} \right) \nn\\
& - \frac13 {{h_{10}^{(f)}}_{r_f}}^B \left( \p_u {h_{10}^{(f)}}_{AB}  + \p_u D_{(A}{h_{10}^{(f)}}_{B) r_f} \right) - {{h_{10}^{(f)}}_{r_f}}^B \left( D_A {h_{10}^{(f)}}_{r_f B} - D_B {h_{10}^{(f)}}_{r_f A} \right) \nn \\
&    + \frac16 \p_u {{h_{11}^{(f)}}_{r_f}}^B \left({h_{1m}^{(f)}}_{AB} + D_B {h_{10}^{(f)}}_{r_f A} - D_A {h_{10}^{(f)}}_{r_f B} \right).  \nn
\end{align}
We can use relations on the unit sphere $R_{ABCD}=\bg_{AC}\bg_{BD}-\bg_{AD}\bg_{BC}$ to simplify the expression of $ {h_{21}}_{uA}$. More precisely, one can verify that
\be
\begin{split}
&D^2 \left( D_A {h_{10}^{(f)}}_{r_f B} - D_B {h_{10}^{(f)}}_{r_f A} \right) \\
=&2D^C \left( D_A D_{(C} {h_{10}^{(f)}}_{B) r_f } - D_B D_{(C} {h_{10}^{(f)}}_{A) r_f } \right) + 2\left( D_B {h_{10}^{(f)}}_{r_f A} - D_A {h_{10}^{(f)}}_{r_f B} \right).
\end{split}
\ee
We define
\be
\bar C_{1 AB}={h_{1m}^{(f)}}_{AB} + 2 D_{(A} {h_{10}^{(f)}}_{B) r_f },
\ee
which can be understood as the traceful shear at order $G$ without supertranslation, i.e., the bared traceful shear tensor at order $G$. Then, the angular momentum aspect can be given in a more compact form as
\begin{align}
{N_2}_A\equiv&{h_{21}}_{uA}\nn\\
=&  {h_{21}^{(f)}}_{u_f A}  + D_A {h_{22}^{(f)}}_{u_f r_f}+ \frac13 \p_u {h_{22}^{(f)}}_{r_f A} + \frac13 \p_u D_A {h_{23}^{(f)}}_{r_f r_f} \nn
 \\
&  +\beta_1 \p_u {N_1}_{A} + \frac23 D_A \beta_1 m_1  + \frac13 D_A \beta_1 \left( {h_{11}^{(f)}}_{u_f u_f}   + \p_u {h_{12}^{(f)}}_{r_f r_f}\right) \nn \\
& + \frac16  D^B \beta_1 D_A D_B  {h_{12}^{(f)}}_{r_f r_f}
 + \frac12 (D^B \beta_1 -  {{h_{10}^{(f)}}_{r_f}}^B ) D^C \left( D_A \bar C_{1B C} - D_B \bar C_{1A C} \right) \nn\\
&  + \frac13 (D^B \beta_1  -  {{h_{10}^{(f)}}_{r_f}}^B ) \p_u \left(  {h_{10}^{(f)}}_{AB}  +  D_{(A}{h_{10}^{(f)}}_{B) r_f}\right)  \\
&    - \frac13 {h_{10}^{(f)}}_{r_f A} \p_u {h_{12}^{(f)}}_{r_f r_f} 
  - \frac{1}{6} {h_{10}^{(f)}}_{r_f B} D_A D^B \p_u {h_{12}^{(f)}}_{r_f r_f}+ \frac{1}{12} {h_{1m}^{(f)}}_{A B} D^B \p_u {h_{12}^{(f)}}_{r_f r_f} \nn \\ 
&  + \frac16 \p_u {h_{12}^{(f)}}_{r_f r_f} \left(D_B D_A {{h_{10}^{(f)}}_{r_f}}^B  - D^2  {h_{10}^{(f)}}_{r_f A} + D_A {{h_{1m}^{(f)}}_{B_f}}^{B_f} - D^B {h_{1m}^{(f)}}_{A B}\right) \nn \\
&  + \frac{1}{12} \left(D^B  {h_{10}^{(f)}}_{r_f A} D_B \p_u {h_{12}^{(f)}}_{r_f r_f} - D_A  {{h_{10}^{(f)}}_{r_f}}^B D_B \p_u {h_{12}^{(f)}}_{r_f r_f} \right) \nn \\
&    + \frac16 \p_u {{h_{11}^{(f)}}_{r_f}}^B \left({h_{1m}^{(f)}}_{AB} + D_B {h_{10}^{(f)}}_{r_f A} - D_A {h_{10}^{(f)}}_{r_f B} \right),  \nn
\end{align}
where $m_1$ and ${N_1}_A$ are the order $G$ Bondi mass and angular momentum aspects, respectively. The first line in the expression for the angular momentum aspect at this order comes from the order $G^2$ metric in flat Bondi coordinates. These terms involve precisely the same kinds of quantities that appear in the linear order angular momentum aspect. Notably, the order $G$ supertranslation $\beta_1$ enters repeatedly and plays a prominent role in the full expression. In addition, the contributions inherited from the linear order metric remain substantial. The smoothness condition at this order is ${h_{21}^{(f)}}_{r_f r_f}=0$. The news tensor at order $G^2$
\be
{N_2}_{AB}=\p_u {C_2}_{AB}=\p_{u_f} {h_{2m}^{(f)}}_{A  B }  - \frac12 \bg_{AB} \bg^{CD} \p_{u_f} {h_{2m}^{(f)}}_{C  D } ,
\ee
is completely fixed by the traceless part of the order $G^2$ transverse metric in the flat Bondi coordinates, applying the constraints in \eqref{constraints}. The order $G^2$ mass aspect has a very compact form and is given by
\be
m_2={h_{21}^{(f)}}_{u_f u_f}  + 2 \p_u {h_{22}^{(f)}}_{u_f r_f}  + \frac12 \p_u^2 {h_{23}^{(f)}}_{r_f r_f}  + \beta_1 \p_u m_1.
\ee
The contribution from the order $G^2$ metric in the flat Bondi coordinates contains exactly the same type of structures that appear in the linear order mass aspect. The supertranslation dependence at this order is governed by the time evolution of the linear order mass aspect.


\section{Concluding remarks}

In this paper, we obtain the generic coordinate transformation in asymptotic expansions that relates the harmonic gauge and the NU gauge up to the second PM order. This allows us to identify the physical data at null infinity, including the shear tensor, the mass aspect, and the angular momentum aspect. In particular, their explicit dependence on supertranslation is specified. We expect that the concrete relations between harmonic and NU gauge established here could be an important stepping stone for the future investigation about some puzzling facts at 2PM. A notable example is the puzzle of angular-momentum loss in gravitational scattering, where the leading contribution appears either at $\cO(G^2)$ or $\cO(G^3)$ depending on different gauge choices under the supertranslation at $\cO(G)$ \cite{Damour:2020tta,Jakobsen:2021smu,Mougiakakos:2021ckm,Herrmann:2021lqe,Herrmann:2021tct,Bini:2021gat,Riva:2021vnj,Veneziano:2022zwh,Manohar:2022dea,DiVecchia:2022owy,Bini:2022enm,Bini:2022wrq,Heissenberg:2022tsn,Heissenberg:2023uvo,Heissenberg:2024umh,Mao:2024ryq}. The $\cO(G^2)$ angular momentum aspect obtained in the present work may shed light on this discrepancy and clarify the structure of angular-momentum flux in the PM expansion. A particularly meaningful test can be performed in the context of gravitational bremsstrahlung, where the 2PM metric in harmonic gauge satisfying the Einstein equation is well established \cite{Kovacs:1978eu,Bel:1981be,Jakobsen:2021smu}.


\section*{Acknowledgments}

This work is supported in part by the National Natural Science Foundation of China (NSFC) under Grants No.~12475059 and No.~11935009, and by Tianjin University Self-Innovation Fund Extreme Basic Research Project Grant No.~2025XJ21-0007.


\begin{thebibliography}{10}

\bibitem{LIGOScientific:2016aoc}
{\bfseries LIGO Scientific, Virgo} Collaboration, B.~P. Abbott {\em et~al.}, ``{Observation of Gravitational Waves from a Binary Black Hole Merger},'' \href{http://dx.doi.org/10.1103/PhysRevLett.116.061102}{{\em Phys. Rev. Lett.} {\bfseries 116} no.~6, (2016) 061102}, \href{http://arxiv.org/abs/1602.03837}{{\ttfamily arXiv:1602.03837 [gr-qc]}}.

\bibitem{Einstein:1916cc}
A.~Einstein, ``{Approximative Integration of the Field Equations of Gravitation},''
{\em Sitzungsber. Preuss. Akad. Wiss. Berlin (Math. Phys.)} {\bfseries 1916} (1916) 688--696.

\bibitem{Einstein:1918btx}
A.~Einstein, ``{\"{U}ber Gravitationswellen},''
{\em Sitzungsber. Preuss. Akad. Wiss. Berlin (Math. Phys.)} {\bfseries 1918} (1918) 154--167.

\bibitem{Bondi:1962px}
H.~Bondi, M.~G.~J. van~der Burg, and A.~W.~K. Metzner, ``{Gravitational waves in general relativity. 7. Waves from axisymmetric isolated systems},'' \href{http://dx.doi.org/10.1098/rspa.1962.0161}{{\em Proc. Roy. Soc. Lond. A} {\bfseries 269} (1962) 21--52}.

\bibitem{Sachs:1962wk}
R.~K. Sachs, ``{Gravitational waves in general relativity. 8. Waves in asymptotically flat space-times},'' \href{http://dx.doi.org/10.1098/rspa.1962.0206}{{\em Proc. Roy. Soc. Lond. A} {\bfseries 270} (1962) 103--126}.

\bibitem{Madler:2016xju}
T.~M{\"a}dler and J.~Winicour, ``{Bondi-Sachs Formalism},'' \href{http://dx.doi.org/10.4249/scholarpedia.33528}{{\em Scholarpedia} {\bfseries 11} (2016) 33528}, \href{http://arxiv.org/abs/1609.01731}{{\ttfamily arXiv:1609.01731 [gr-qc]}}.

\bibitem{Newman:1961qr}
E.~Newman and R.~Penrose, ``{An Approach to gravitational radiation by a method of spin coefficients},''
\href{http://dx.doi.org/10.1063/1.1724257}{{\em J. Math. Phys.} {\bfseries 3} (1962) 566--578}.

\bibitem{Newman:1962cia}
E.~T. Newman and T.~W.~J. Unti, ``{Behavior of Asymptotically Flat Empty Spaces},'' \href{http://dx.doi.org/10.1063/1.1724303}{{\em J. Math. Phys.} {\bfseries 3} no.~5, (1962) 891}.

\bibitem{Barnich:2011ty}
G.~Barnich and P.-H. Lambert, ``{A Note on the Newman-Unti group and the BMS charge algebra in terms of Newman-Penrose coefficients},'' \href{http://dx.doi.org/10.1155/2012/197385}{{\em Adv. Math. Phys.} {\bfseries 2012} (2012) 197385}, \href{http://arxiv.org/abs/1102.0589}{{\ttfamily arXiv:1102.0589 [gr-qc]}}.

\bibitem{Newman:1965ik}
E.~T. Newman and R.~Penrose, ``{10 exact gravitationally-conserved quantities},'' \href{http://dx.doi.org/10.1103/PhysRevLett.15.231}{{\em Phys. Rev. Lett.} {\bfseries 15} (1965) 231--233}.

\bibitem{Newman:1968uj}
E.~T. Newman and R.~Penrose, ``{New conservation laws for zero rest-mass fields in asymptotically flat space-time},'' \href{http://dx.doi.org/10.1098/rspa.1968.0112}{{\em Proc. Roy. Soc. Lond. A} {\bfseries 305} (1968) 175--204}.

\bibitem{Winicour}
J.~Winicour, ``{Some Total Invariants of Asymptotically Flat Space‐Times},'' \href{http://dx.doi.org/10.1063/1.1664652}{{\em J. Math. Phys.} {\bfseries 9} (1968) 861--867}.

\bibitem{Geroch:1977big}
R.~Geroch, \href{http://dx.doi.org/10.1007/978-1-4684-2343-3_1}{``{Asymptotic Structure of Space-Time},''} in {\em {Symposium on Asymptotic Structure of Space-Time}}.
\newblock 1977.

\bibitem{Prior}
C.~R. {Prior}, ``{Angular Momentum in General Relativity. I. Definition and Asymptotic Behaviour},'' \href{http://dx.doi.org/10.1098/rspa.1977.0073}{{\em Proc. Roy. Soc. Lond. A} {\bfseries 354} (1977) 379--405}.

\bibitem{Ashtekar:1978zz}
A.~Ashtekar and R.~O. Hansen, ``{A unified treatment of null and spatial infinity in general relativity. I - Universal structure, asymptotic symmetries, and conserved quantities at spatial infinity},'' \href{http://dx.doi.org/10.1063/1.523863}{{\em J. Math. Phys.} {\bfseries 19} (1978) 1542--1566}.

\bibitem{Ashtekar:1979xeo}
A.~Ashtekar and A.~Magnon-Ashtekar, ``{Energy-Momentum in General Relativity},'' \href{http://dx.doi.org/10.1103/PhysRevLett.43.181}{{\em Phys. Rev. Lett.} {\bfseries 43} no.~3, (1979) 181}.

\bibitem{Ashtekar:1979iaf}
A.~Ashtekar and M.~Streubel, ``{On angular momentum of stationary gravitating systems},'' \href{http://dx.doi.org/10.1063/1.524242}{{\em J. Math. Phys.} {\bfseries 20} no.~7, (1979) 1362--1365}.

\bibitem{Ashtekar:1981bq}
A.~Ashtekar and M.~Streubel, ``{Symplectic Geometry of Radiative Modes and Conserved Quantities at Null Infinity},'' \href{http://dx.doi.org/10.1098/rspa.1981.0109}{{\em Proc. Roy. Soc. Lond. A} {\bfseries 376} (1981) 585--607}.

\bibitem{Geroch:1981ut}
R.~P. Geroch and J.~Winicour, ``{Linkages in general relativity},'' \href{http://dx.doi.org/10.1063/1.524987}{{\em J. Math. Phys.} {\bfseries 22} (1981) 803--812}.

\bibitem{Dray:1984rfa}
T.~Dray and M.~Streubel, ``{Angular momentum at null infinity},'' \href{http://dx.doi.org/10.1088/0264-9381/1/1/005}{{\em Class. Quant. Grav.} {\bfseries 1} no.~1, (1984) 15--26}.

\bibitem{Barnich:2010eb}
G.~Barnich and C.~Troessaert, ``{Aspects of the BMS/CFT correspondence},'' \href{http://dx.doi.org/10.1007/JHEP05(2010)062}{{\em JHEP} {\bfseries 05} (2010) 062}, \href{http://arxiv.org/abs/1001.1541}{{\ttfamily arXiv:1001.1541 [hep-th]}}.

\bibitem{Barnich:2011mi}
G.~Barnich and C.~Troessaert, ``{BMS charge algebra},'' \href{http://dx.doi.org/10.1007/JHEP12(2011)105}{{\em JHEP} {\bfseries 12} (2011) 105}, \href{http://arxiv.org/abs/1106.0213}{{\ttfamily arXiv:1106.0213 [hep-th]}}.

\bibitem{Flanagan:2015pxa}
E.~E. Flanagan and D.~A. Nichols, ``{Conserved charges of the extended Bondi-Metzner-Sachs algebra},'' \href{http://dx.doi.org/10.1103/PhysRevD.95.044002}{{\em Phys. Rev. D} {\bfseries 95} no.~4, (2017) 044002}, \href{http://arxiv.org/abs/1510.03386}{{\ttfamily arXiv:1510.03386 [hep-th]}}. [Erratum: Phys.Rev.D 108, 069902 (2023)].

\bibitem{Blanchet:2020ngx}
L.~Blanchet, G.~Comp\`ere, G.~Faye, R.~Oliveri, and A.~Seraj, ``{Multipole expansion of gravitational waves: from harmonic to Bondi coordinates},'' \href{http://dx.doi.org/10.1007/JHEP02(2021)029}{{\em JHEP} {\bfseries 02} (2021) 029}, \href{http://arxiv.org/abs/2011.10000}{{\ttfamily arXiv:2011.10000 [gr-qc]}}.

\bibitem{Blanchet:2023pce}
L.~Blanchet, G.~Comp\`ere, G.~Faye, R.~Oliveri, and A.~Seraj, ``{Multipole expansion of gravitational waves: memory effects and Bondi aspects},'' \href{http://dx.doi.org/10.1007/JHEP07(2023)123}{{\em JHEP} {\bfseries 07} (2023) 123}, \href{http://arxiv.org/abs/2303.07732}{{\ttfamily arXiv:2303.07732 [gr-qc]}}.

\bibitem{Thorne:1980ru}
K.~S. Thorne, ``{Multipole Expansions of Gravitational Radiation},'' \href{http://dx.doi.org/10.1103/RevModPhys.52.299}{{\em Rev. Mod. Phys.} {\bfseries 52} (1980) 299--339}.

\bibitem{Blanchet:1985sp}
L.~Blanchet and T.~Damour, ``{Radiative gravitational fields in general relativity I. general structure of the field outside the source},'' \href{http://dx.doi.org/10.1098/rsta.1986.0125}{{\em Phil. Trans. Roy. Soc. Lond. A} {\bfseries 320} (1986) 379--430}.

\bibitem{Blanchet:1986dk}
L.~Blanchet, ``{Radiative gravitational fields in general relativity. 2. Asymptotic behaviour at future null infinity},'' \href{http://dx.doi.org/10.1098/rspa.1987.0022}{{\em Proc. Roy. Soc. Lond. A} {\bfseries 409} (1987) 383--399}.

\bibitem{Blanchet:1987wq}
L.~Blanchet and T.~Damour, ``{Tail Transported Temporal Correlations in the Dynamics of a Gravitating System},'' \href{http://dx.doi.org/10.1103/PhysRevD.37.1410}{{\em Phys. Rev. D} {\bfseries 37} (1988) 1410}.

\bibitem{Damour:1990ji}
T.~Damour and B.~R. Iyer, ``{PostNewtonian generation of gravitational waves. 2. The Spin moments},'' {\em Ann. Inst. H. Poincare Phys. Theor.} {\bfseries 54} (1991) 115--164.

\bibitem{Blanchet:1992br}
L.~Blanchet and T.~Damour, ``{Hereditary effects in gravitational radiation},'' \href{http://dx.doi.org/10.1103/PhysRevD.46.4304}{{\em Phys. Rev. D} {\bfseries 46} (1992) 4304--4319}.

\bibitem{Blanchet:1993ng}
L.~Blanchet, ``{Time asymmetric structure of gravitational radiation},'' \href{http://dx.doi.org/10.1103/PhysRevD.47.4392}{{\em Phys. Rev. D} {\bfseries 47} (1993) 4392--4420}.

\bibitem{Blanchet:1996vx}
L.~Blanchet, ``{Gravitational radiation reaction and balance equations to postNewtonian order},'' \href{http://dx.doi.org/10.1103/PhysRevD.55.714}{{\em Phys. Rev. D} {\bfseries 55} (1997) 714--732}, \href{http://arxiv.org/abs/gr-qc/9609049}{{\ttfamily arXiv:gr-qc/9609049}}.

\bibitem{Blanchet:2018yqa}
L.~Blanchet and G.~Faye, ``{Flux-balance equations for linear momentum and center-of-mass position of self-gravitating post-Newtonian systems},'' \href{http://dx.doi.org/10.1088/1361-6382/ab0d4f}{{\em Class. Quant. Grav.} {\bfseries 36} no.~8, (2019) 085003}, \href{http://arxiv.org/abs/1811.08966}{{\ttfamily arXiv:1811.08966 [gr-qc]}}.

\bibitem{Compere:2019gft}
G.~Comp\`ere, R.~Oliveri, and A.~Seraj, ``{The Poincar\'e and BMS flux-balance laws with application to binary systems},'' \href{http://dx.doi.org/10.1007/JHEP10(2020)116}{{\em JHEP} {\bfseries 10} (2020) 116}, \href{http://arxiv.org/abs/1912.03164}{{\ttfamily arXiv:1912.03164 [gr-qc]}}.

\bibitem{Bonga:2018gzr}
B.~Bonga and E.~Poisson, ``{Coulombic contribution to angular momentum flux in general relativity},'' \href{http://dx.doi.org/10.1103/PhysRevD.99.064024}{{\em Phys. Rev. D} {\bfseries 99} no.~6, (2019) 064024}, \href{http://arxiv.org/abs/1808.01288}{{\ttfamily arXiv:1808.01288 [gr-qc]}}.

\bibitem{Blanchet:2013haa}
L.~Blanchet, ``{Gravitational Radiation from Post-Newtonian Sources and Inspiralling Compact Binaries},'' \href{http://dx.doi.org/10.1007/s41114-024-00050-z}{{\em Living Rev. Rel.} {\bfseries 27} (2024) 4}, \href{http://arxiv.org/abs/1310.1528}{{\ttfamily arXiv:1310.1528 [gr-qc]}}.

\bibitem{Duarte:2022mxj}
M.~Duarte, J.~C. Feng, E.~Gasperin, and D.~Hilditch, ``{Peeling in generalized harmonic gauge},'' \href{http://dx.doi.org/10.1088/1361-6382/ac89c5}{{\em Class. Quant. Grav.} {\bfseries 39} no.~21, (2022) 215003}, \href{http://arxiv.org/abs/2205.09405}{{\ttfamily arXiv:2205.09405 [gr-qc]}}.

\bibitem{Satishchandran:2019pyc}
G.~Satishchandran and R.~M. Wald, ``{Asymptotic behavior of massless fields and the memory effect},'' \href{http://dx.doi.org/10.1103/PhysRevD.99.084007}{{\em Phys. Rev. D} {\bfseries 99} no.~8, (2019) 084007}, \href{http://arxiv.org/abs/1901.05942}{{\ttfamily arXiv:1901.05942 [gr-qc]}}.

\bibitem{Mao:2024urq}
P.~Mao and B.~Zeng, ``{Note on post-Minkowskian expansion and Bondi coordinates},'' \href{http://dx.doi.org/10.1016/j.nuclphysb.2025.117111}{{\em Nucl. Phys. B} {\bfseries 1019} (2025) 117111}, \href{http://arxiv.org/abs/2405.11953}{{\ttfamily arXiv:2405.11953 [gr-qc]}}.

\bibitem{Veneziano:2022zwh}
G.~Veneziano and G.~A. Vilkovisky, ``{Angular momentum loss in gravitational scattering, radiation reaction, and the Bondi gauge ambiguity},'' \href{http://dx.doi.org/10.1016/j.physletb.2022.137419}{{\em Phys. Lett. B} {\bfseries 834} (2022) 137419}, \href{http://arxiv.org/abs/2201.11607}{{\ttfamily arXiv:2201.11607 [gr-qc]}}.

\bibitem{Damour:2020tta}
T.~Damour, ``{Radiative contribution to classical gravitational scattering at the third order in $G$},'' \href{http://dx.doi.org/10.1103/PhysRevD.102.124008}{{\em Phys. Rev. D} {\bfseries 102} no.~12, (2020) 124008}, \href{http://arxiv.org/abs/2010.01641}{{\ttfamily arXiv:2010.01641 [gr-qc]}}.

\bibitem{Flanagan:2023jio}
E.~E. Flanagan and D.~A. Nichols, ``{Fully nonlinear transformations of the Weyl-Bondi-Metzner-Sachs asymptotic symmetry group},'' \href{http://dx.doi.org/10.1007/JHEP03(2024)120}{{\em JHEP} {\bfseries 03} (2024) 120}, \href{http://arxiv.org/abs/2311.03130}{{\ttfamily arXiv:2311.03130 [gr-qc]}}.

\bibitem{Jakobsen:2021smu}
G.~U. Jakobsen, G.~Mogull, J.~Plefka, and J.~Steinhoff, ``{Classical Gravitational Bremsstrahlung from a Worldline Quantum Field Theory},'' \href{http://dx.doi.org/10.1103/PhysRevLett.126.201103}{{\em Phys. Rev. Lett.} {\bfseries 126} no.~20, (2021) 201103}, \href{http://arxiv.org/abs/2101.12688}{{\ttfamily arXiv:2101.12688 [gr-qc]}}.

\bibitem{Mougiakakos:2021ckm}
S.~Mougiakakos, M.~M. Riva, and F.~Vernizzi, ``{Gravitational Bremsstrahlung in the post-Minkowskian effective field theory},'' \href{http://dx.doi.org/10.1103/PhysRevD.104.024041}{{\em Phys. Rev. D} {\bfseries 104} no.~2, (2021) 024041}, \href{http://arxiv.org/abs/2102.08339}{{\ttfamily arXiv:2102.08339 [gr-qc]}}.

\bibitem{Herrmann:2021lqe}
E.~Herrmann, J.~Parra-Martinez, M.~S. Ruf, and M.~Zeng, ``{Gravitational Bremsstrahlung from Reverse Unitarity},'' \href{http://dx.doi.org/10.1103/PhysRevLett.126.201602}{{\em Phys. Rev. Lett.} {\bfseries 126} no.~20, (2021) 201602}, \href{http://arxiv.org/abs/2101.07255}{{\ttfamily arXiv:2101.07255 [hep-th]}}.

\bibitem{Herrmann:2021tct}
E.~Herrmann, J.~Parra-Martinez, M.~S. Ruf, and M.~Zeng, ``{Radiative classical gravitational observables at $ \mathcal{O} $(G$^{3}$) from scattering amplitudes},'' \href{http://dx.doi.org/10.1007/JHEP10(2021)148}{{\em JHEP} {\bfseries 10} (2021) 148}, \href{http://arxiv.org/abs/2104.03957}{{\ttfamily arXiv:2104.03957 [hep-th]}}.

\bibitem{Bini:2021gat}
D.~Bini, T.~Damour, and A.~Geralico, ``{Radiative contributions to gravitational scattering},'' \href{http://dx.doi.org/10.1103/PhysRevD.104.084031}{{\em Phys. Rev. D} {\bfseries 104} no.~8, (2021) 084031}, \href{http://arxiv.org/abs/2107.08896}{{\ttfamily arXiv:2107.08896 [gr-qc]}}.

\bibitem{Riva:2021vnj}
M.~M. Riva and F.~Vernizzi, ``{Radiated momentum in the post-Minkowskian worldline approach via reverse unitarity},'' \href{http://dx.doi.org/10.1007/JHEP11(2021)228}{{\em JHEP} {\bfseries 11} (2021) 228}, \href{http://arxiv.org/abs/2110.10140}{{\ttfamily arXiv:2110.10140 [hep-th]}}.

\bibitem{Manohar:2022dea}
A.~V. Manohar, A.~K. Ridgway, and C.-H. Shen, ``{Radiated Angular Momentum and Dissipative Effects in Classical Scattering},'' \href{http://dx.doi.org/10.1103/PhysRevLett.129.121601}{{\em Phys. Rev. Lett.} {\bfseries 129} no.~12, (2022) 121601}, \href{http://arxiv.org/abs/2203.04283}{{\ttfamily arXiv:2203.04283 [hep-th]}}.

\bibitem{DiVecchia:2022owy}
P.~Di~Vecchia, C.~Heissenberg, and R.~Russo, ``{Angular momentum of zero-frequency gravitons},'' \href{http://dx.doi.org/10.1007/JHEP08(2022)172}{{\em JHEP} {\bfseries 08} (2022) 172}, \href{http://arxiv.org/abs/2203.11915}{{\ttfamily arXiv:2203.11915 [hep-th]}}.

\bibitem{Bini:2022enm}
D.~Bini, T.~Damour, and A.~Geralico, ``{Radiated momentum and radiation reaction in gravitational two-body scattering including time-asymmetric effects},'' \href{http://dx.doi.org/10.1103/PhysRevD.107.024012}{{\em Phys. Rev. D} {\bfseries 107} no.~2, (2023) 024012}, \href{http://arxiv.org/abs/2210.07165}{{\ttfamily arXiv:2210.07165 [gr-qc]}}.

\bibitem{Bini:2022wrq}
D.~Bini and T.~Damour, ``{Radiation-reaction and angular momentum loss at the second post-Minkowskian order},'' \href{http://dx.doi.org/10.1103/PhysRevD.106.124049}{{\em Phys. Rev. D} {\bfseries 106} no.~12, (2022) 124049}, \href{http://arxiv.org/abs/2211.06340}{{\ttfamily arXiv:2211.06340 [gr-qc]}}.

\bibitem{Heissenberg:2022tsn}
C.~Heissenberg, ``{Angular Momentum Loss due to Tidal Effects in the Post-Minkowskian Expansion},'' \href{http://dx.doi.org/10.1103/PhysRevLett.131.011603}{{\em Phys. Rev. Lett.} {\bfseries 131} no.~1, (2023) 011603}, \href{http://arxiv.org/abs/2210.15689}{{\ttfamily arXiv:2210.15689 [hep-th]}}.

\bibitem{Heissenberg:2023uvo}
C.~Heissenberg, ``{Angular momentum loss due to spin-orbit effects in the post-Minkowskian expansion},'' \href{http://dx.doi.org/10.1103/PhysRevD.108.106003}{{\em Phys. Rev. D} {\bfseries 108} no.~10, (2023) 106003}, \href{http://arxiv.org/abs/2308.11470}{{\ttfamily arXiv:2308.11470 [hep-th]}}.

\bibitem{Heissenberg:2024umh}
C.~Heissenberg and R.~Russo, ``{Revisiting gravitational angular momentum and mass dipole losses in the eikonal framework},'' \href{http://dx.doi.org/10.1088/1361-6382/adaabc}{{\em Class. Quant. Grav.} {\bfseries 42} no.~4, (2025) 045014}, \href{http://arxiv.org/abs/2406.03937}{{\ttfamily arXiv:2406.03937 [gr-qc]}}.

\bibitem{Mao:2024ryq}
P.~Mao and B.~Zeng, ``{Supertranslation ambiguity in post-Minkowskian expansion},'' \href{http://dx.doi.org/10.1103/PhysRevD.111.L021502}{{\em Phys. Rev. D} {\bfseries 111} no.~2, (2025) L021502}, \href{http://arxiv.org/abs/2406.07943}{{\ttfamily arXiv:2406.07943 [gr-qc]}}.

\bibitem{Kovacs:1978eu}
S.~J. Kovacs and K.~S. Thorne, ``{The Generation of Gravitational Waves. 4. Bremsstrahlung},'' \href{http://dx.doi.org/10.1086/156350}{{\em Astrophys. J.} {\bfseries 224} (1978) 62--85}.

\bibitem{Bel:1981be}
L.~Bel, T.~Damour, N.~Deruelle, J.~Ibanez, and J.~Martin, ``{Poincar\'e-invariant gravitational field and equations of motion of two pointlike objects: The postlinear approximation of general relativity},'' \href{http://dx.doi.org/10.1007/BF00756073}{{\em Gen. Rel. Grav.} {\bfseries 13} (1981) 963--1004}.

\end{thebibliography}
\providecommand{\href}[2]{#2}\begingroup\raggedright\endgroup

\end{document}